\documentclass{article} 
\usepackage{iclr2019_conference,times}

\usepackage{amsmath,amsfonts,bm}

\def\eqref#1{equation~\ref{#1}}

\def\1{\bm{1}}

\def\vb{{\bm{b}}}

\def\vh{{\bm{h}}}

\def\vv{{\bm{v}}}
\def\vw{{\bm{w}}}

\def\mA{{\bm{A}}}

\def\mD{{\bm{D}}}

\def\mH{{\bm{H}}}

\def\mK{{\bm{K}}}

\def\mQ{{\bm{Q}}}

\def\mV{{\bm{V}}}
\def\mW{{\bm{W}}}

\DeclareMathAlphabet{\mathsfit}{\encodingdefault}{\sfdefault}{m}{sl}
\SetMathAlphabet{\mathsfit}{bold}{\encodingdefault}{\sfdefault}{bx}{n}

\newcommand{\R}{\mathbb{R}}

\usepackage{hyperref}
\usepackage{url}
\usepackage{amsmath}
\usepackage{graphicx}
\usepackage[utf8]{inputenc}
\usepackage{siunitx}
\usepackage{hyperref}
\usepackage{subfig}

\graphicspath{ {./extra/} }

\title{Universal Transforming Geometric Network}

\author{Jin Li \thanks{ Worked performed at ShanghaiTech, School of Information Science and Technology.} \\
University of Chicago\\
Chicago, IL 60637, USA \\
\texttt{jinli11@uchicago.edu} 
}

\iclrfinalcopy 
\begin{document}

\maketitle

\begin{abstract}
The recurrent geometric network (RGN), the first end-to-end differentiable neural architecture for protein structure prediction, is a competitive alternative to existing models. However, the RGN's use of recurrent neural networks (RNNs) as internal representations results in long training time and unstable gradients. And because of its sequential nature, it is less effective at learning global dependencies among amino acids than existing transformer architectures. We propose the Universal Transforming Geometric Network (UTGN), an end-to-end differentiable model that uses the encoder portion of the Universal Transformer architecture as an alternative for internal representations. Our experiments show that compared to RGN, UTGN achieve a $1.7$ \si{\angstrom} improvement on the free modeling portion and a $0.7$ \si{\angstrom} improvement on the template based modeling of the CASP12 competition.
\end{abstract}

\section{Introduction}

Proteins are chains of chemical units, called amino acids, that fold to form three dimensional structures. The ability to predict a protein's structure from its amino acid sequence remain to be the most elusive yet rewarding challenges in computational biology. The structure determines the protein's function, which can be used to help understand life threatening diseases and accelerate drug discover (\cite{Kuntz1992StructurebasedSF}). However, experimental methods for solving a protein's structure is both time consuming and costly, and they only account for a small percentage of known protein sequences.

Earlier computational methods for predicting protein structure includes molecular dynamics (MD), which use physic based equations to simulate the trajectory of a protein's folding process into a stable final 3D conformation (\cite{dMarxAbInitio}). However, this method is computationally expensive and ineffective for larger proteins. Other approaches include using co-evolutional information to predict the residue-residue contact map, which can be used to guide structure prediction methods. With the help of deep learning architectures like convolutional neural networks, contact prediction remains to be the prevailing methods in structure prediction (\cite{Wang2016AccurateDN}). However, because these method does not provide an explicit mapping from sequence to structure, they lack the ability to capture intrinsic information between the sequence and structure.
 
RGNs solve that issue, as it is an end-to-end differentiable model that jointly optimizes the relationships between protein sequences and structure. However, because it uses RNNs as internal representations, training can be both difficult and time consuming (\cite{AlQuraishi2019EndtoEndDL}).

In this paper, we propose a modification to the RGN architecture. Inspired by the recent successes of the transformer models in the NLP community, we replace the LSTMs in the RGN model with the encoder portion of the Universal Transformer (UT) as the internal representation (\cite{Dehghani2019UniversalT}). By doing so, the model is faster to train and it is contextually informed by all subsequent symbols. As a result, it is better at learning global dependencies among the amino acids than RNNs.
 
\begin{figure}[t]
\begin{center}
\includegraphics[width=139mm,scale=0.5]{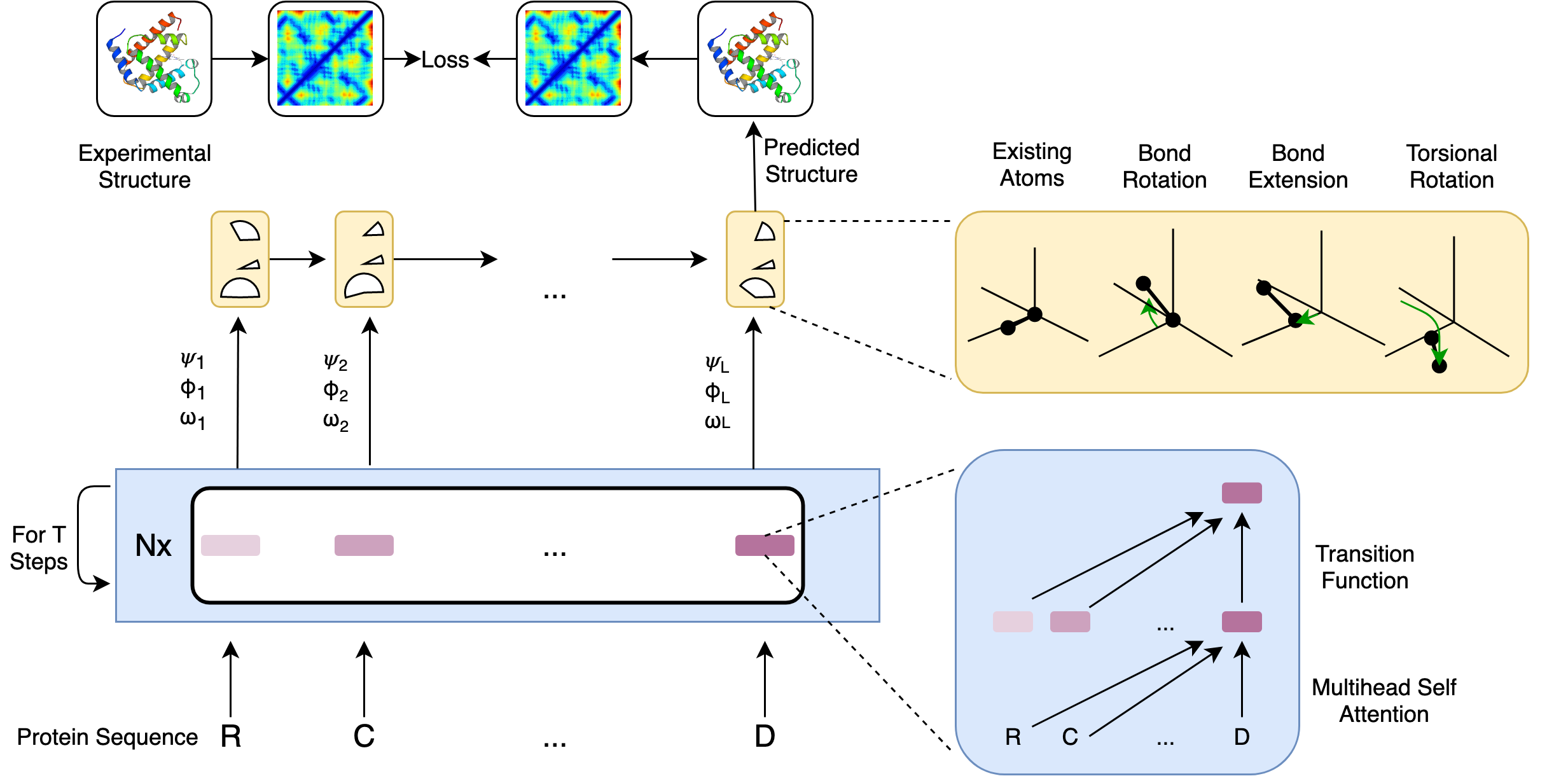}
\end{center}
\caption{Vector representation of the amino acids are fed into the encoder of the UT. During each time step, the encoder applies a multi-head self attention mechanism and then a transition function to incorporate information from all previous states. In addition, the ACT mechanism decides which representation remain static. During each time step, the UT iteratively refines the internal representations of the inputs until a maximum time step is reached or until ACT mechanism halts every representation. Next, the UTGN constructs 3 torsional angle from each internal representation, which is used to create the 3D Cartesian coordinates for the protein structure. Then the loss function dRMSD is calculated by creating the distance map for the experimental structure and the predicted structure and calculating the distances between those two distance map. Back-propagation is used to optimize the weights of the architecture.}
\label{fig:UT_Full}
\end{figure}

The UTGN operates by first taking a sequence of vector representation of the amino acids and applying the universal transformer architecture to iteratively refine a sequence of internal representations. Next, it uses the internal states to construct three torsional angles for each position, which is used to construct the 3D Cartesian structure (Figure \ref{fig:UT_Full}).
 
Our experiments show that UTGN achieve an improvement of $1.7 \si{\angstrom}$ in RMSD and $0.013$ in TM-Score for the free modeling portion of CASP12. In addition, the UTGN achieved an improvement of $0.7 \si{\angstrom}$ in RMSD and $0.008$ in TM-Score for the template based modeling portion.

\section{Model Description}

\subsection{Input Representation}

We represent each amino acid in the protein sequence of size $L$ as a $20$ dimensional one-hot encoding. Next, we derive the Multiple Sequence Alignment (MSA) from JackHMMer and use it to calculate the $L \times 20$ Position-Specific Scoring Matrix (PSSM) (\cite{Potter2018HMMERWS}). Then, we normalize the PSSM values to between 0 and 1 and concatenate it with the one-hot encoding. After feeding this into a fully connected layer of dimension $d$, we add positional encodings as in \cite{Vaswani2017AttentionIA} as follows

\begin{equation}
PE_{(j, 2i)} = \sin(\frac{j}{10000^{\frac{2i}{d}}} )   
\end{equation}

\begin{equation}
PE_{(j, 2i + 1)} = \cos(\frac{j}{10000^{\frac{2i}{d}}} )
\end{equation}

where $j$ is the position of the vector in the protein sequence and $i$ is the index of that vector.

\subsection{Universal Transformer}

We use the multi-layer encoder portion of the Universal Transformer (UT) for internal representation. This neural architecture operates by recurring over representations of each of the positions of the input sequences. Unlike recurrent neural networks, which recur over positions in the sequence, UT recurs over revisions of the vector representations of each position (\cite{Dehghani2019UniversalT}). In each time step, the representations are revised by passing through $N$ layers, where each layer consists of a self-attention mechanism to exchange information across all positions in the sequence in the previous representation, followed by a transition function.

More specifically, given an input sequence of length $L$, we initialize a matrix $ \mH^{0} \in \R^{L \times d}$. Each new representation $ \mH^{t} \in \R^{L \times d} $ at time step $t$ is determined by first applying the multi-head dot-product self-attention (\cite{Vaswani2017AttentionIA}) mechanism. We compute the scaled dot-product attention using queries $\mQ$, keys $\mK$, and values $\mV$ as follows

\begin{equation}
    \mathrm{ATTENTION}(\mQ, \mK, \mV) = \mathrm{SOFTMAX}(\frac{\mQ\mK^{T}}{\sqrt{d}})\mV
\end{equation}

For each head $i$, we map state $\mH^{t}$ to queries, keys, and values using learned matrices $\mW_{i}^{Q} \in  \R^{d \times \frac{d}{k}}$ , $\mW_{i}^{K} \in  \R^{d \times \frac{d}{k}}$ , and $\mW_{i}^{V} \in  \R^{d \times \frac{d}{k}}$, where $k$ is the number of heads (\cite{Vaswani2017AttentionIA}). Next, we apply the scaled-dot product attention to each head, concatenate them, and multiply the result by a learned matrix $\mW^{O} \in  \R^{d \times d}$. 

\begin{equation}
    \mathrm{MULTIHEAD}(\mH^{t}) = \mathrm{CONCAT}(\mathrm{head_{1}, ..., \mathrm{head}_{k}})\mW^{O}
\end{equation}

\begin{equation}
    \mathrm{head}_{i} = \mathrm{ATTENTION}(\mH^{t}\mW_{i}^{Q}, \mH^{t}\mW_{i}^{K}, \mH^{t}\mW_{i}^{V})
\end{equation}

We pass through the first layer of the encoder as follows

\begin{equation}
    \mH_{1}^{t} = \mathrm{LAYER}_{1}(\mH^{t - 1})
\end{equation}

where 

\begin{equation}
    \mathrm{LAYER}_{p}(\mH) = \mathrm{LAYERNORM}(\mA^{t} + \mathrm{TRANSITION}(\mA^{t}))
\end{equation}

\begin{equation}
    \mA^{t} = \mathrm{LAYERNORM}(\mH + \mathrm{MULTIHEAD}(\mH))
\end{equation}

where LAYERNORM is defined in \cite{Ba2016LayerN} and TRANSITION is either a one dimensional separable convolution (\cite{Chollet2016XceptionDL}) or a fully-connected layer. In addition, each layer has different weights and $p$ indicates the layer number. Between the multi-head attention and transition function, we incorporate both residual connections and dropout (\cite{Srivastava2014DropoutAS}). 

For an encoder with $N$ layers, we have 

\begin{equation}
    \mH^{t} = \mH_{N}^{t} = \mathrm{LAYER}_{N}(\mH_{N - 1}^{t})
\end{equation}

\begin{figure}[h]
\begin{center}
\includegraphics[width=120mm,scale=0.5]{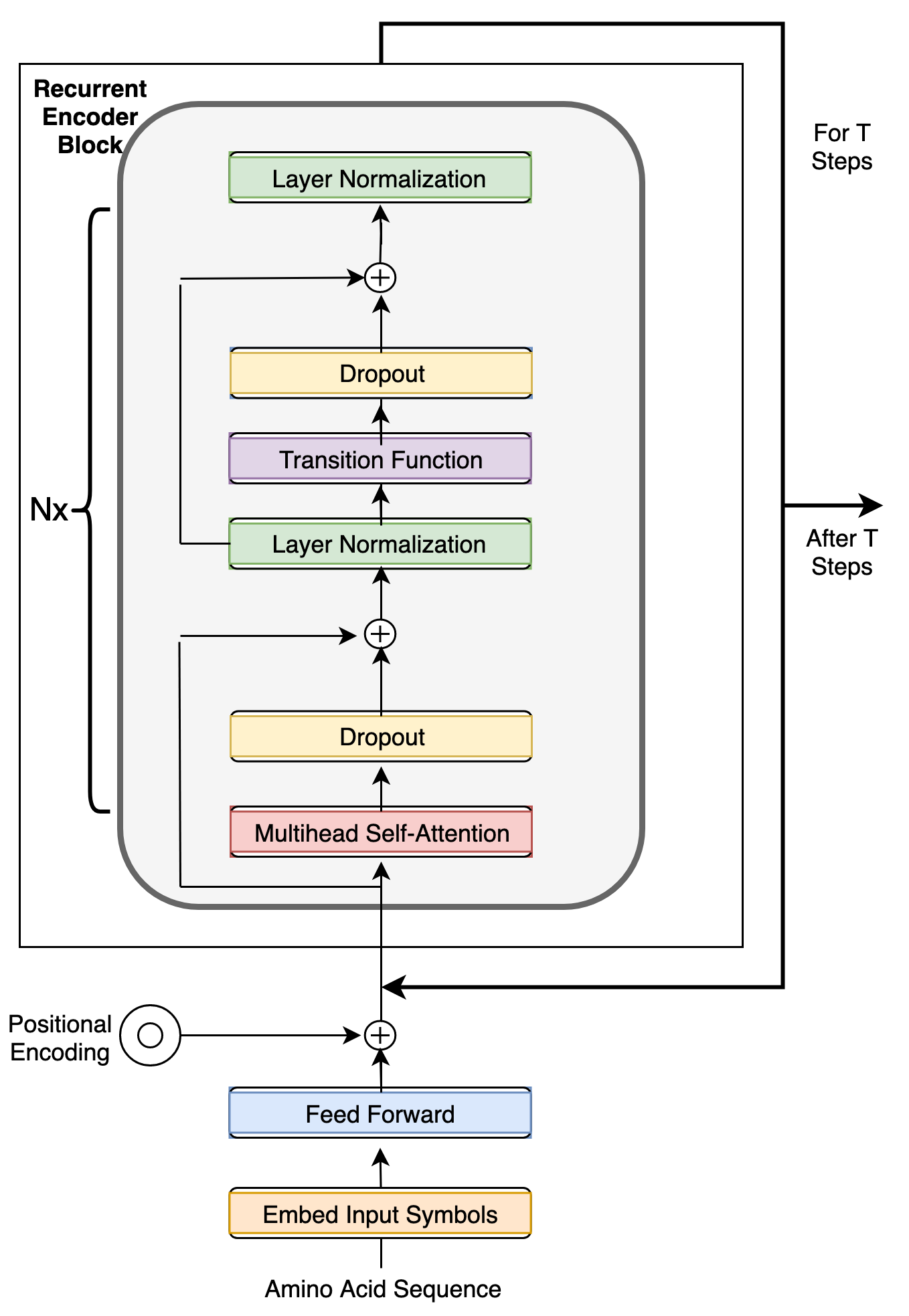}
\end{center}
\caption{Complete architecture of the encoder portion of the universal transformer.}
\label{fig:UT}
\end{figure}

In contrast to the original UT model, we do not add positional encodings on each time step; rather, the positional encoding is only added at the initial starting phase (see Figure \ref{fig:UT} for a complete model).

\subsection{Dynamic Halting}

Because we wish to expend more computing resources on amino acids with ambiguous relevancy, we use the Adaptive Computation Time (ACT) to dynamically halt changes in certain representations (\cite{Graves2016AdaptiveCT}). For each step and each symbol, if the scalar halting probability predicted by the model exceeds a threshold, the state representation is simply copied to the next time step. Recurrence continues until all representations are halted or the maximum number of steps are met. 

\subsection{Structure Construction}

As in \cite{AlQuraishi2019EndtoEndDL}, we use the final states for each position to construct the three torsional angles $\varphi$, $\psi$, $ \omega$. These angles represent the geometry of the protein spanned by the backbone atoms $N$, $C^{\alpha}$, $C'$. Though bond lengths and angles also vary, their variation is limited enough that we can assume them to be fixed. We will also ignore the side chains of the protein, as our focus is on the backbone atoms. The resulting angles at each position is then translated into the 3D coordinates for the backbone.

More specifically, at position $j$, the corresponding angle triplet $\phi_{j} = (\psi_{j}, \varphi_{j}, \omega_{j})$ is calculated as follows

\begin{equation}
    \phi_{j} = arg(p_{j}exp(i\Phi))
\end{equation}

\begin{equation}
    p_{j} = \mathrm{SOFTMAX}(\mW_{\phi}\vh_{j} + \vb_{\phi})
\end{equation}

where $\vh_{j}$ is the $j^{\mathrm{th}}$ row of matrix $\mH^{T}$, $ \mW_{\phi}$, $\vb_{\phi}$, $\Phi$ are learned weights, and $arg$ is the complex valued argument function. In addition, $\Phi$ defines an alphabet of size $m$ whose letters correspond to triplets of torsional angles over the 3-torus. Next, recurrent geometric units convert the sequence of torsional angles $ (\phi_{1}, ..., \phi_{L})$ into 3D Cartesian coordinates as follows

\begin{equation} 
\Tilde{c_{k}} = r_{k \: \mathrm{mod} \: 3} 
    \begin{pmatrix} 
    \cos(\theta_{k \: \mathrm{mod} \: 3}) \\
    \cos(\phi_{\left \lfloor{\frac{k}{3}}\right \rfloor, k \: \mathrm{mod} \: 3 }) sin(\theta_{k \: \mathrm{mod} \: 3}) \\
    \sin(\phi_{\left \lfloor{\frac{k}{3}}\right \rfloor, k \: \mathrm{mod} \: 3 }) sin(\theta_{k \: \mathrm{mod} \: 3})
    \end{pmatrix}
\end{equation}

$$
    m_{k} = c_{k - 1} - c_{k - 2}
$$
$$
    n_{k} = m_{k - 1} \times  \widehat{m}_{k}
$$
$$
    M_{k} = 
    \begin{pmatrix}
        \widehat{m}_{k}, \widehat{n}_{k} \times \widehat{m}_{k}, \widehat{n}_{k}
    \end{pmatrix}
$$
$$
c_{k} = M_{k}\Tilde{c_{k}} + c_{k - 1}
$$

where $r_{k}$ is the length of the bond connecting atoms $k - 1$ and $k$, $\theta$ is the bond angle formed by atoms $k - 2$, $k - 1$, and $k$, $\phi_{\left \lfloor{\frac{k}{3}}\right \rfloor, k \: \mathrm{mod} \: 3 }$ is the predicted torsional angle formed by atoms $k - 2$ and $k - 1$, $\widehat{m}$ is the unit-normalized version of $m$, $\times$ is the cross product, and $c_{k}$ is the position of the newly predicted atom $k$. The sequence $(c_{1}, ... c_{3L})$ form the final Cartesian coordinates of the  protein backbone chain structure.

For training, the weights are optimized through the dRMSD loss function between the predicted and expected coordinates. This computes the pairwise distances between each atom in either the predicted or expected structure, and then finds the distance between those distances. More specifically, 

\begin{equation}
    \tilde{d_{j,k}} = \| c_{j} - c_{k} \|_{2}
\end{equation}
$$
    d_{j,k} = \tilde{d_{j,k}}^{(exp)} - \tilde{d_{j,k}}^{(pred)}
$$
$$
    \mathrm{dRMSD} = \frac{\| \mD \|_{2}}{L(L - 1)}
$$

where $d_{j,k}$ are elements of matrix $\mD$. We chose this loss function because it is differentiable and captures both local and global aspects of the protein structure.

\section{Experiments and Analysis}

\subsection{Training Data and Batching}
We evaluate our models with the CASP12 ProteinNet dataset with a thinning of 90\%, which consists of around 50,000 structures (\cite{AlQuraishi2019ProteinNetAS}). The train and validation set contains all sequences and structures that exist prior to the CASP12 competition. The test set is the targets of CASP12, which consists of both the template-based modeling (TBM), intended to assess the prediction of targets with structural homologs in the Protein Data Bank, and the free modeling (FM), intended to test a model's ability to predict novel structures (\cite{Moult1995ALE}). 
In the train and validation set, entries with missing residues were annotated and are not included in the calculation of dRMSD. Sequences with similar lengths are batched together with a batch size of 32.

\subsection{Model Parameters}
The dimension of the feed forward layer that connected the input to the UT encoder was $256$. We use $8$ heads and $6$ layers for the UT encoder architecture. The ACT threshold is $0.5$ and the maximum number of ACT recurrence was $10$. If a feed forward layer was used for the transition function (UTGN-FF), the feed forward dimension was $128$. If a separable convolution is used instead (UTGN-SepConv), the kernel size is set to $3$ and the stride was set to $1$. In addition, we set the alphabet size to 60 for the angularization layer. The UTGN architecture amounts to about $2$ million trainable parameters. For point of comparison, we train the RGN model with a size of $240$, which is also around $2$ million trainable parameters.

\subsection{Optimizer}
We used the ADAM optimizer with $\beta_{1} = 0.95$, $\beta_{2} = 0.99$, and learning rate of $0.001$ (\cite{Kingma2014AdamAM}). In addition, the loss function for optimization was length normalized (dRMSD / protein length).

\subsection{Regularization}
We apply a dropout probability of $0.10$ in the UT encoder architecture (\cite{Srivastava2014DropoutAS}). In addition, gradients are clipped using norm re-scaling with a threshold of 5.0 (\cite{Pascanu2012UnderstandingTE}). Furthermore, we perform early stopping when the validation loss failed to change noticeably in $10$ epochs.

\subsection{Analysis}

We evaluate our model using two metrics: root mean squared deviation (RMSD) and Template Modelling (TM) Score (\cite{Zhang2004ScoringFF}). RMSD is
calculated by

\begin{equation}
    \mathrm{RMSD}(\vv, \vw) = \sqrt{\frac{1}{n}\sum_{i = 1}^{n}\| v_{i} - w_{i}\|^{2} }
\end{equation}

where $\vv$ and $\vw$ are two sets of $n$ points. This metric has the advantage that it does not require two structures to be globally aligned, and is able to detect regions of high agreement even if the global structure is not aligned. However, RMSD is very sensitive to protein length, leading to higher RMSD for longer proteins.  The TM Score is calculated by

\begin{equation}
    \mathrm{TMScore}(\vv, \vw) = \sum_{i}\frac{1}{1 + (\frac{D_{i}}{D_{0}})^{2} }
\end{equation}

where $D_{i} = \| \vv - \vw \|$. TM scores are length normalized, and take values between $0$ and $1$, with higher values indicating better alignment. A TM score $ <0.17$ corresponds to random alignment whereas a TM score of $>0.5$ correspond to the same protein fold (\cite{Xu2010HowSI}).

After evaluating our model, we found that the UTGN with separable convolution as the transition function performed better than the RGN model in the free modeling category by $9$ percent in the RMSD metric and by $7$ percent in the TM score metric (Table 2). For the template based modeling, the improvement was $4$ percent for RMSD and $4$ percent for TM score (Table 1).

From training both the RGN and UTGN model, we note that RGNs tend to suffer from heavily from exploding gradients, whereas the UTGN model never had that issue. In addition, the RGN model takes around $6$ times longer for each epoch, and the UTGN model converged to its result about $2$ times faster. Furthermore, we found that UTGNs have more stable initializations, whereas different initializations in RGNs can produce very different evaluation results. 

Because RGNs and UTGNs must learn very deep neural networks from scratch and do not include any biophysical priors into the model, training a state-of-the-art model would require months of training and $10$ times more parameters. Though we only train for a few days and with only $2$ million parameters, we show that UTGNs have the potential to outperform RGNs.

\begin{table}[h]
\begin{center}
\begin{tabular}{lll}
\multicolumn{1}{c}{\bf Model}  &\multicolumn{1}{c}{\bf dRMSD (\si{\angstrom})} &\multicolumn{1}{c}{\bf TM score}
\\ \hline \\
RGN            &17.8 &0.200\\
UTGN-FF        &17.6  &0.198\\
UTGN-SepConv   &\textbf{17.1}  &\textbf{0.208}\\
\end{tabular}
\caption{The average dRMSD (lower is better) and TM score (higher is better) achieved by RGN and UTGN models in the TBM category for CASP12.}
\end{center}
\label{table:TBM}
\end{table}

\begin{table}[h]
\begin{center}
\begin{tabular}{lll}
\multicolumn{1}{c}{\bf Model}  &\multicolumn{1}{c}{\bf dRMSD (\si{\angstrom})} &\multicolumn{1}{c}{\bf TM score}
\\ \hline \\
RGN            &19.8  &0.181\\
UTGN-FF        &19.4  &0.174\\
UTGN-SepConv   &\textbf{18.1}  &\textbf{0.194}\\
\end{tabular}
\caption{The average dRMSD (lower is better) and TM score (higher is better) achieved by RGN and UTGN models in the FM category for CASP12.}
\end{center}
\label{table:FM}
\end{table}

\section{Discussion}

Before RGN was introduced, the protein prediction competition was dominated by complex models that fuse together multiple pipelines (\cite{Yang2014TheIS}). They tend to incorporate biological priors, like co-evolutionary information and secondary structure, that significantly improved their model performance. But the RGN model show to be a very competitive option without biological priors, outperforming the CASP11 model in the free modeling category (\cite{AlQuraishi2019EndtoEndDL}). Just like end to end differentiable models were able to replace complex pipelines in image recognition, we expect end-to-end differentiable architectures like UTGN to eventually replace the complex pipeline in protein structure prediction. 

The biggest bottlenecks for training RGNs is time. The recurrent neural network portion is unstable, leading to gradient explosions. In addition, different initializations produce very different model performance, requiring researchers to try many different initializations. Furthermore, a fully refined model may take months to train (\cite{AlQuraishi2019EndtoEndDL}). As a result, it may take a significant amount of time to search for optimal parameters. UTGNs, however, are able to solve many of these problems, as replacing RNNs with transformers lead to a more stable and easily parallelizable model. 

UTGNs are also better at learning global dependencies than RGNs. In RNNs, as the length of the path between two amino acids increase, information flow decreases. In contrast, the UT effectively has a global receptive field, as each new representation is contextually informed by all previous representations.

Some possible extensions for UTGN include using pre-trained embedding representations of amino acids like UniRep (\cite{Alley2019UnifiedRP}). This can  replace the need to calculate PSSMs for each new sequence, which further reduces the prediction time. In addition, instead of using static positional encodings, we could train relative position representations along with the transformer (\cite{Shaw2018SelfAttentionWR}). Or we could incorporate more information in the input sequence, like secondary structure predictions.

\section{Conclusion}

This paper introduces UTGN, an end-to-end protein structure prediction architecture that uses a universal transformer as an internal representation. As opposed to the existing RGN model, UTGN is better at learning relationships of long range dependencies in the amino acids. In addition, the UTGN perform slightly better, converge much quicker, and is more stable to train. This progress shows that end-to-end differentiable protein prediction architectures can become competitive models in the protein folding problem.

The code for UTGN is available at: \url{https://github.com/JinLi711/3DProteinPrediction}

\subsubsection*{Acknowledgments}

We are grateful for Jie Zheng and Suwen Zhao for providing insightful guidance and commentary. We are also thankful for ShanghaiTech, School of Information Science and Technology for providing access to its computer cluster.

\clearpage
\bibliography{iclr2019_conference}
\bibliographystyle{iclr2019_conference}

\clearpage
\appendix
\section*{Appendices}

\section{Universal Transformer Architecture}

\clearpage
\section{CASP12 Comparison}

\begin{figure}[h]
\begin{center}

\subfloat{{
\includegraphics[width=60mm,scale=0.5]{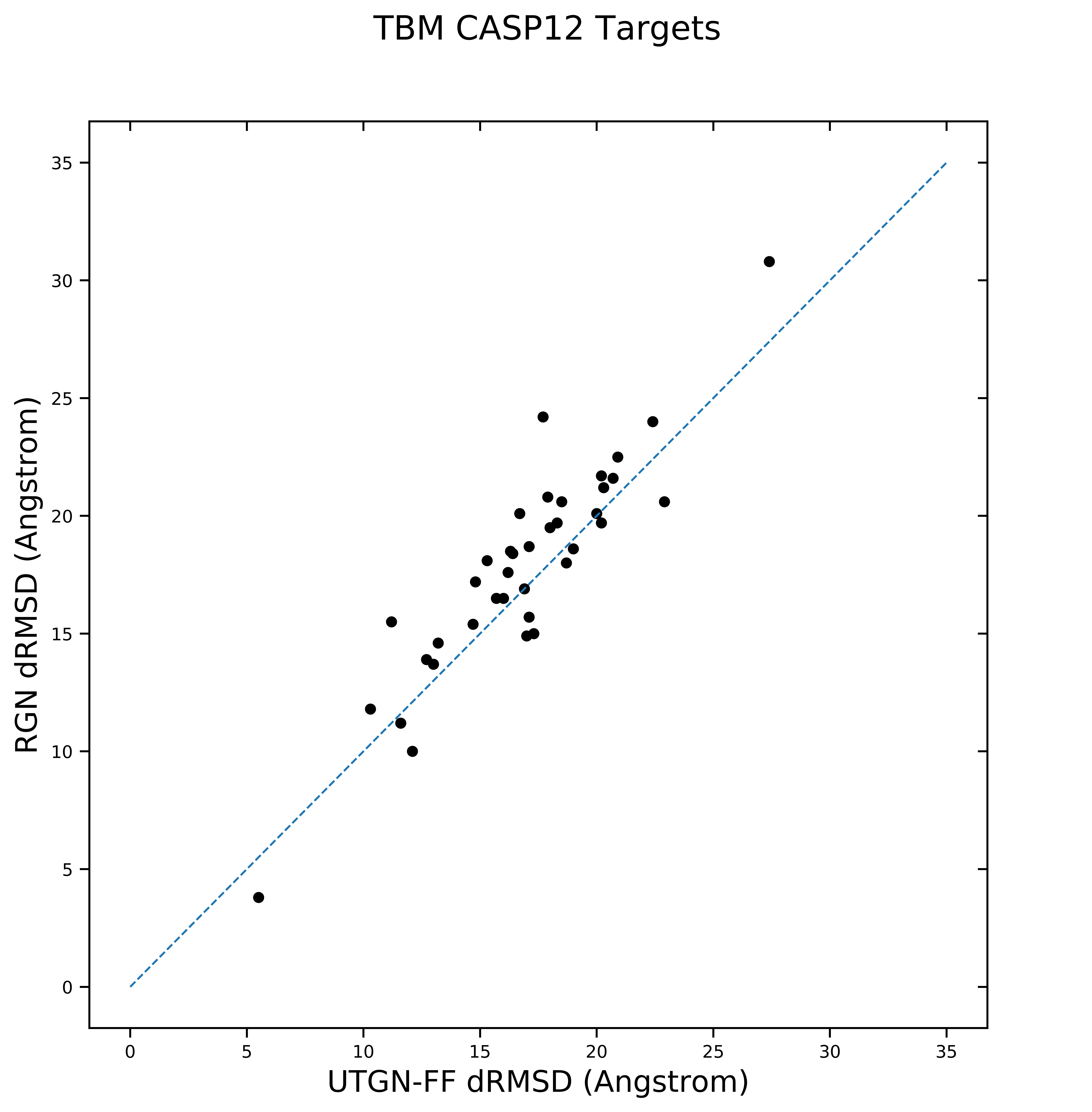}
}}%
\subfloat{{
\includegraphics[width=60mm,scale=0.5]{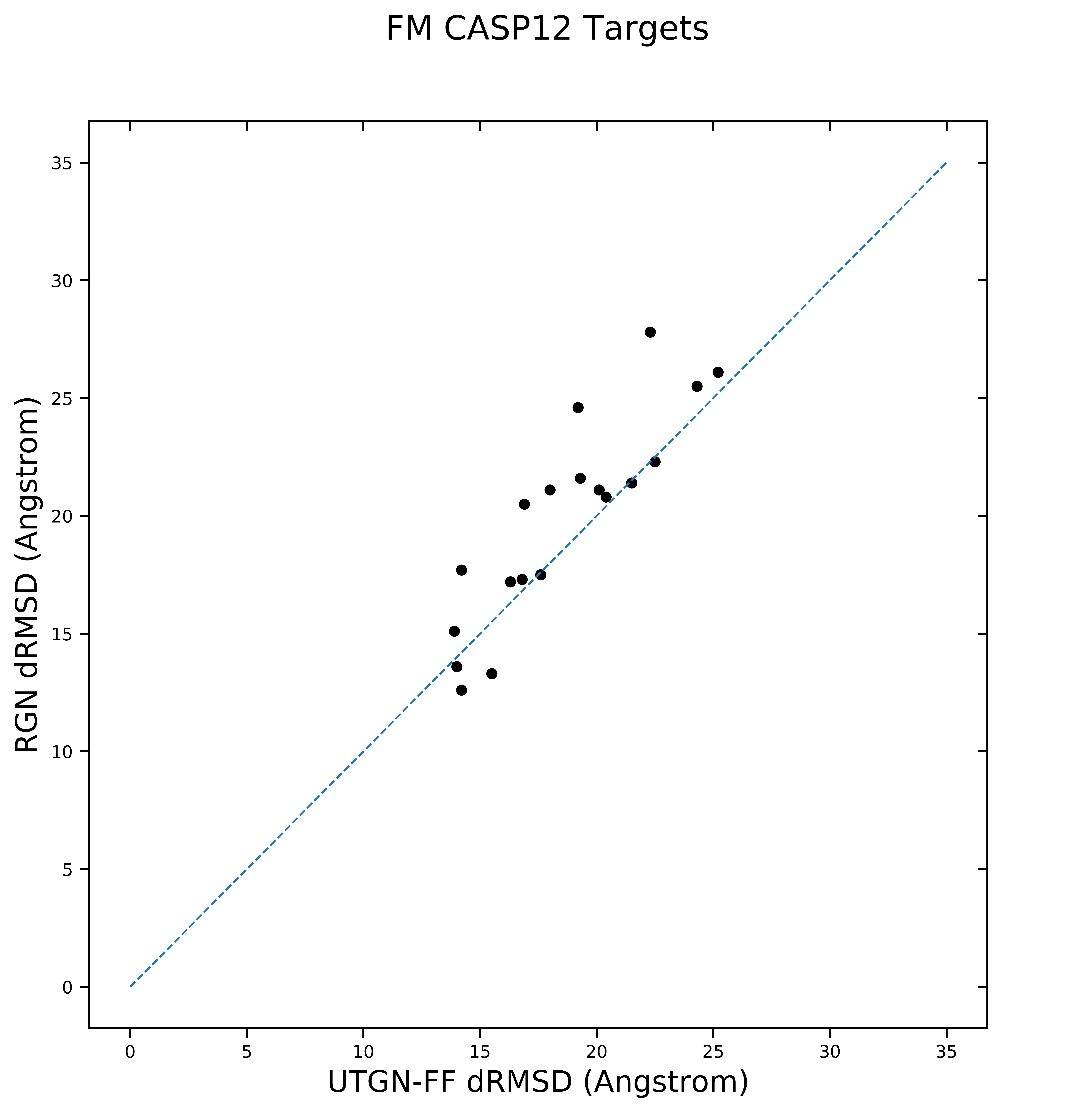}
}}

\end{center}
\caption{Scatter-plot comparing individual FM and TBM predictions of RGN and UTGN feed forward. Points above the blue line indicates better UTGN performance.}
\label{plot:FF}
\end{figure}

\begin{figure}[h]
\begin{center}

\subfloat{{
\includegraphics[width=60mm,scale=0.5]{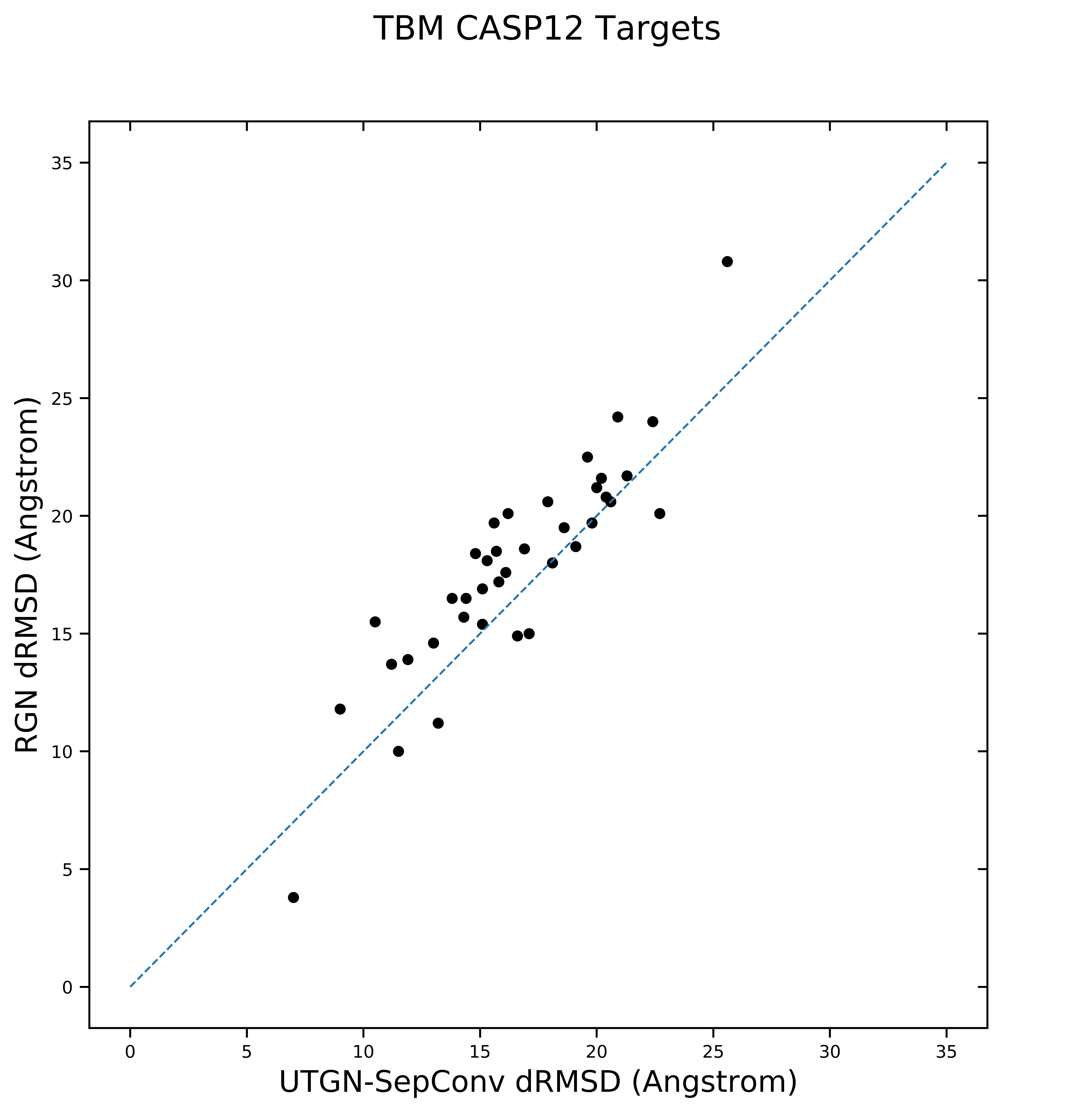}
}}%
\subfloat{{
\includegraphics[width=60mm,scale=0.5]{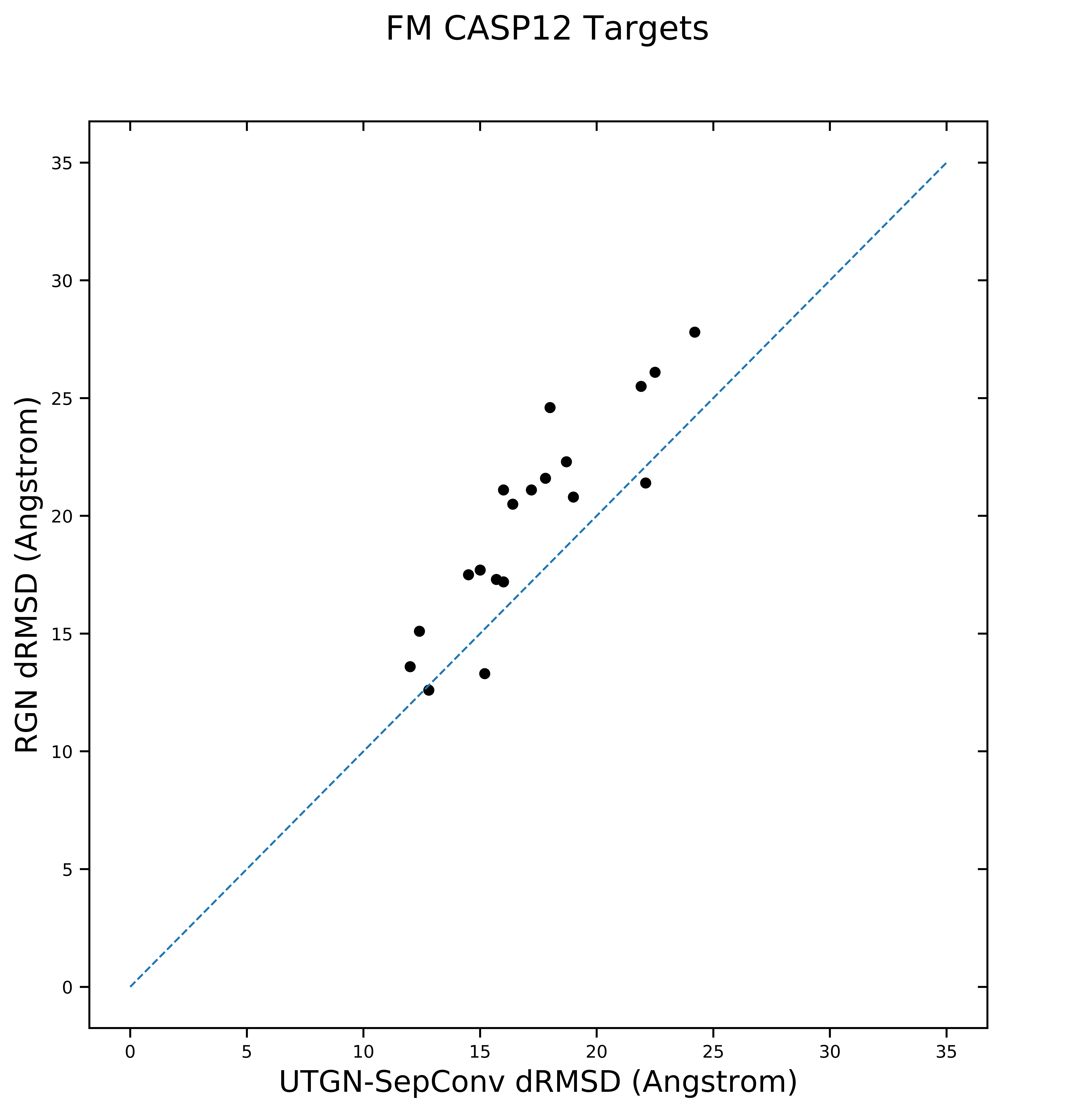}
}}

\end{center}
\caption{Scatter-plot comparing individual FM and TBM predictions of RGN and UTGN with separable convolution. Points above the blue line indicates better UTGN performance.}
\label{plot:Sep}
\end{figure}

\clearpage
\section{CASP12 Samples}
\begin{figure}[h]
\begin{center}
\includegraphics[width=139mm,scale=0.5]{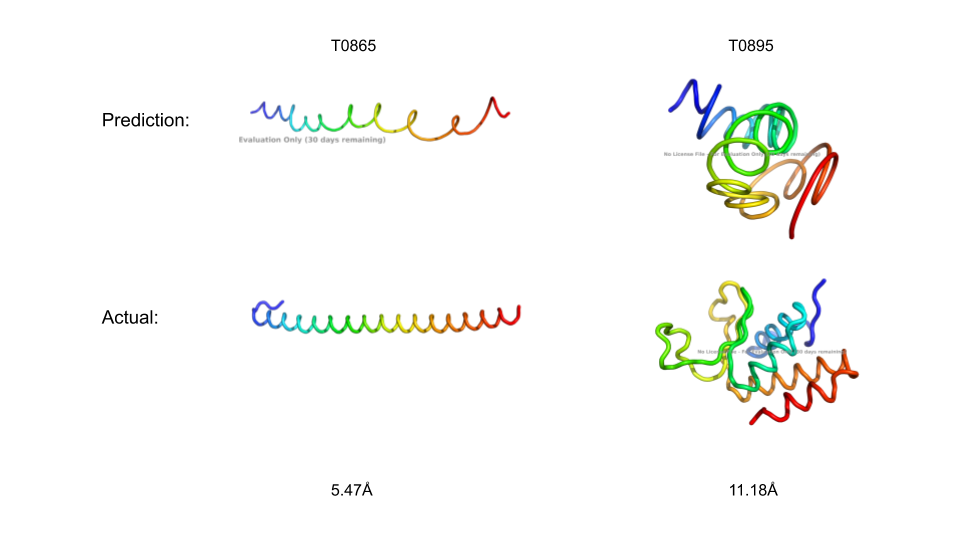}
\end{center}
\caption{Comparison between predicted and actual structure of proteins T0865 (RMSD of $5.47 \si{\angstrom}$) and T0895 (RMSD of $11.18 \si{\angstrom}$) from CASP12 competition.}
\label{fig:comparison}
\end{figure}

\end{document}